%% file: hadron2011.tex
\documentclass[a4paper,11pt]{article}

\usepackage{contribution}

\input{econfmacros}
\input{contributionmacros}

\include{definitions}

\begin{document}

\input{hadron2011contribution}

\end{document}

%% file: econfmacros.tex
\newcommand{\weblink}[2][]{%
    \ifthenelse{\equal{#1}{}}%
    {\textnormal{\url{#2}}}%
    {\textnormal{\href{#2}{#1}}}%
}

\def\beq{\begin{equation}}
\def\eeq#1{\label{#1}\end{equation}}
\def\eeqn{\end{equation}}

\def\beqa{\begin{eqnarray}}
\def\eeqa#1{\label{#1}\end{eqnarray}}
\def\eeqan{\end{eqnarray}}

\let\bar=\overbar

\def\bra#1{\left\langle{ #1} \right|}
\def\ket#1{\left| {#1} \right\rangle}

\def\Dslash{\not{\hbox{\kern-4pt $D$}}}
\def\dslash{\not{\hbox{\kern-2pt $\del$}}}

\def\ee{e^+e^-}

\def\msb{{\bar{\ssstyle M \kern -1pt S}}}

%% file: contributionmacros.tex
\newcommand{\contribution}[7][]{%
  \clearpage
  \thispagestyle{plain}
  \ifthenelse{\equal{#1}{}}
  {\hypersetup{pdftitle={#2}}}
  {\hypersetup{pdftitle={#1}}}
  \hypersetup{pdfauthor={{#3} {#4}}}
  {\centering\normalfont\LARGE\bfseries\sffamily #2 \par\nobreak}
  \lhead{}
  \chead{%
    \textit{\footnotesize XIV International Conference on Hadron Spectroscopy
      (\weblink[\textit{hadron2011}]{http://www.hadron2011.de}), 13-17 June 2011, Munich, Germany}%
  }
  \rhead{}
  \bigskip
  \begin{center}
    {#3} {#4}\ifthenelse{\equal{#6}{}}{}{\footnote{\weblink[#6]{mailto:#6}}}
    \ifthenelse{\equal{#7}{}}{}{#7} \\
    \textit{#5}
  \end{center}
  \bigskip
}

\renewcommand{\abstract}[1]{%
  \begin{center}
    \begin{minipage}{0.85\textwidth}
      \begin{footnotesize}
        #1
      \end{footnotesize}
    \end{minipage}
  \end{center}
  \bigskip
}

%% file: definitions.tex
\def\evhh{\ev{\tfrac{1}{2}\tfrac{1}{2}}}
\def\evso{\ev{\mathbf{L}\cdot\mathbf{S}} }
\def\evt{\ev{\mathbf T}}

\newcommand{\rf}[1]{(\ref{#1})}

\def\etab{\end{tabular}}  
  
\def\bit{\begin{itemize}}
\def\eit{\end{itemize}}

\def\bml{\begin{multline}}
\def\eml{\end{multline}}
\def\be{\begin{equation}} 
\def\ee{\end{equation}}  
\def\bea{\begin{eqnarray}}  
\def\eea{\end{eqnarray}} 
\def\bes{\begin{equation*}} 
\def\ees{\end{equation*}}  
\def\beas{\begin{eqnarray*}}  
\def\eeas{\end{eqnarray*}} 
\def\bmu{\begin{multline}}
\def\emu{\end{multline}}
\def\bal{\begin{array}{l}} 	
\def\eal{\end{array}}

\def\MeV{\textrm{ MeV}}

\def\letterS{S}
\def\letterP{P}
\def\letterD{D}
\def\letterF{F}
\def\greek#1{\def\letter{#1}
\ifx\letter\letterS\Sigma
\else
	\ifx\letter\letterP\Pi
	\else
		\ifx\letter\letterD\Delta
		\else
			\ifx\letter\letterF\Phi
			\else XXX
			\fi
		\fi
	\fi
\fi
}

\def\cc{c\overline c}

\def\cc{c\overline c}

\def\bb{b\overline b}

\def\bb{b\overline b}

\newcommand{\ev}[1]{\langle #1 \rangle}

\renewcommand{\u}[1]{\textrm{#1}}
\def\cn#1#2#3{^{#1}\u {#2}_{#3}}     
\def\an{\cn} 

\newcommand{\uP}{\textrm{P}}

\def\xQN#1,{\def\Element{#1}%
\ifx\Element\endpiece
\else 
	\if\Element/\def\ApplyOrLiteral{1} 
	\else
		\if\ApplyOrLiteral0\ElementRule\Element 
		\else\Element\def\ApplyOrLiteral{0}
		\fi
	\fi
	\expandafter\xQN
\fi}

\def\xNQ#1,{\def\ElementRule{#1}%
\ifx\ElementRule\endpiece
\else 
	\if\ElementRule/\def\ApplyOrLiteral{1} 
	\else
		\if\ApplyOrLiteral0\ElementRule\Element 
		\else\ElementRule\def\ApplyOrLiteral{0}
		\fi
	\fi
	\expandafter\xNQ
\fi}

\def\k|#1>{\ket{#1}} 
\def\b<#1|{\bra{#1}} 

\def\bk<#1|#2>{\langle #1 | #2 \rangle}
\def\Bk#1<#2|#3>{\def\ElementRule{#1}\def\ApplyOrLiteral{0}
\xQN/,\langle,#2,xxx,\vert #3 \rangle}
\def\bK#1<#2|#3>{Undefined}

\def\BK#1<#2|#3>{\def\ElementRule{#1}\def\ApplyOrLiteral{0}
\xQN/,\langle,#2,/,|,#3,/,\rangle,xxx,}
\def\KB<#1|#2>#3{\def\Element{#3}\def\ApplyOrLiteral{0}
\xNQ/,\langle,#1,/,|,#2,/,\rangle,xxx,}

\def\me<#1|#2|#3>{\langle #1\vert #2\vert #3\rangle}
\def\ME#1<#2|#3|#4>{\def\ElementRule{#1}\def\ApplyOrLiteral{0}
\xQN/,\langle,#2,/,\vert,/,#3,/,\vert,#4,/,\rangle,xxx,}
\def\EM<#1|#2|#3>#4{\def\Element{#4}\def\ApplyOrLiteral{0}
\xNQ/,\langle,#1,/,\vert,/,#2,/,\vert,#3,/,\rangle,xxx,}

\def\rme<#1|#2|#3>{\langle #1\Vert #2\Vert #3\rangle}
\def\RME#1<#2|#3|#4>{\def\ElementRule{#1}\def\ApplyOrLiteral{0}
\xQN/,\langle,#2,/,\Vert,/,#3,/,\Vert,#4,/,\rangle,xxx,}
\def\EMR<#1|#2|#3>#4{\def\Element{#4}\def\ApplyOrLiteral{0}
\xNQ/,\langle,#1,/,\Vert,/,#2,/,\Vert,#3,/,\rangle,xxx,}

\def\endpiece{xxx}



%% file: hadron2011contribution.tex
%
%
%
%
%
{  


%

\contribution[P-wave spin-spin splitting and meson loops]  
{P-wave spin-spin splitting and meson loops}  
{T. J.}{Burns}  
{INFN Roma\\ Piazzale A. Moro 2\\ Roma, 00185\\ ITALY}  
{Timothy.Burns@roma1.infn.it}  
{}
%

\abstract{%
In quark potential models the hyperfine splitting of P-wave mesons is zero in the nonrelativistic limit, a prediction strikingly confirmed by experiment in both charmonia and bottomonia. The result, however, ignores the coupling of bare quarkonia to meson-meson pairs. This coupling causes mass shifts among the states and so could potentially spoil the quark model prediction. This turns out not to be the case: in a variety of models the hyperfine splitting remains small despite large mass shifts. This is shown to be a generic feature of models in which the coupling involves the creation of a light quark pair with spin-one and the quark spin wavefunctions are conserved. This talk reports on the results of ref. \cite{Burns:2011fu}. 
}
%


In quark potential models the mass $M_{SLJ}$ of a meson of spin $S$, orbital angular momentum $L$ and total angular momentum $J$ can be expressed in perturbation theory, 
\be
M_{SLJ}	=M+\Delta_{s}\evhh_S
+\Delta_{t}\evt_{SLJ}
+\Delta_{o}\evso_{SLJ},
\label{mass}
\ee
in terms of expectations values $M$, $\Delta_{s}$, $\Delta_{t}$ and $\Delta_{o}$ of the spin-independent, spin-spin, tensor and spin-orbit terms. For P-wave mesons the hyperfine splitting $\Delta_s$, which can be expressed in terms of the meson masses by taking the appropriate linear combination of the above,
\be
\frac{1}{9}\left(M_{\an 3P0}+3M_{\an 3P1}+5M_{\an 3P2}\right)-M_{\an 1\uP1}=\Delta_{s},\label{zerohyperfine}
\ee
is zero in the nonrelativistic limit. The experimental charmonia \cite{Dobbs:2008ec} and bottomonia \cite{:2011zp,Adachi:2011ji} masses are in excellent agreement with this prediction :
\beas
\overline M_{\chi_c(1\uP)}-M_{h_c(1\uP)} &=&+0.02\pm 0.19\pm 0.13 \MeV \textrm{(CLEO)},\\
\overline M_{\chi_b(1\uP)}-M_{h_b(1\uP)} &=& +2\pm 4\pm 1 \MeV \textrm{(BaBar)},\\
\overline M_{\chi_b(1\uP)}-M_{h_b(1\uP)} &=& +1.62\pm 1.52 \MeV \textrm{(Belle)},\\
\overline M_{\chi_b(2\uP)}-M_{h_b(2\uP)} &=& +0.48^{+1.57}_{-1.22} \MeV \textrm{(Belle)}.
\eeas

The quark model result ignores the effect of the coupling of bare quarkonia to meson-meson pairs. This ``unquenching'' causes mass shifts, and since the $\chi_0$, $\chi_1$, $\chi_2$ and $h$ have different spin and total angular momenta, their couplings and therefore mass shifts differ. This leads to deviations from the quenched mass formula \rf{mass}, which one might expect could spoil the quark model result.

Remarkably, this is not the case. Table \ref{shiftstable} shows the mass shifts $\Delta M_{SLJ}$ of P-wave $\cc$ and $\bb$ due to coupling  to pseudoscalar and vector mesons, computed in a variety of different approaches   \footnote{The values quoted for ref. \cite{Liu:2011yp} correct a factor of 2 in the coupling of $\chi_{b0}$ to bottom-strange meson pairs.}. Although the mass shifts can be large, the relative shift between any two states is much smaller,  which to some extent explains the empirical success of quenched quark models \cite{Ono:1983rd,Barnes:2007xu}. The relative shifts are, however, still large compared to the experimental hyperfine splittings. It is therefore striking to note that their linear combination
\be
-\frac{1}{9}\left(\Delta M_{\an 3P0}+3\Delta M_{\an 3P1}+5\Delta M_{\an 3P2}\right)+\Delta M_{\an 1\uP1},
\label{indhyp}
\ee
which is the correction to equation \rf{zerohyperfine} due to unquenching, is much smaller still: these induced hyperfine splittings are presented in the final column ``Ind.'' of the table. It thus appears that there is some mechanism in place protecting the smallness of the hyperfine splitting. This is particularly interesting given that the various models differ in several respects.

\begin{table}
\begin{center}
\begin{tabular}{ll D{.}{.}{-1}		D{.}{.}{-1} 		D{.}{.}{-1}		D{.}{.}{-1}		D{.}{.}{-1}}
\hline
Ref.	&	 &\multicolumn{1}{c}{$\Delta M_{\an 3P0}$}	
					&\multicolumn{1}{c}{$\Delta M_{\an 3P1}$}
								&\multicolumn{1}{c}{$\Delta M_{\an 3P2}$}
											&\multicolumn{1}{c}{$\Delta M_{\an 1\uP1}$}
														&\multicolumn{1}{c}{$\textrm{Ind.}$}\\
\hline
\cite{Barnes:2007xu}&1\uP, $\cc$	&459			&496			&521			&504			&-1.8			\\
\cite{Kalashnikova:2005ui}&1\uP, $\cc$	&198			&215			&228			&219			&-1.3			\\
\cite{Li:2009ad}&1\uP, $\cc$&35			&38			&63			&52			&-2.9			\\	
\cite{Yang:2010am}&1\uP, $\cc$&131			&152			&175			&162			&-0.4			\\
\cite{Ono:1983rd}&1\uP, $\cc$	&173			&180			&185			&182			&0.0			\\
\cite{Ono:1983rd}&1\uP, $\bb$	&43			&44			&45			&44			&-0.4			\\
\cite{Ono:1983rd}&2\uP, $\bb$	&55			&56			&58			&57			&0.0			\\
\cite{Liu:2011yp}&1\uP, $\bb$ 	&80.777			&84.823			&87.388			&85.785			&-0.013			\\
\cite{Liu:2011yp}&2\uP, $\bb$	&73.578			&77.608			&80.146			&78.522			&-0.048\\
\hline
\end{tabular}
\end{center}
\caption{The magnitudes of the mass shifts computed in various models. The final column ``Ind.'' shows the induced hyperfine splitting due to loop effects.}
\label{shiftstable}
\end{table}

A feature common to all of the models is that the coupling involves the creation of a light quark pair in spin triplet. The quark spin and spatial degrees of freedom factorise so that the amplitude for the coupling can be expressed as a linear combination of spatial matrix elements weighted by angular momentum recoupling factors. For the coupling to a pair of S-wave mesons there is a single spatial matrix element $A_l$ for each partial wave $l$ \cite{Burns:2007hk}. The corresponding recoupling coefficients $C_{SLJ}^{s_1s_2l}$, for the coupling of a state with $S$, $L$ and $J$ quantum numbers to a pair of S-wave mesons with spins $s_1$, $s_2$, can be deduced from the general expression of ref. \cite{Burns:2006rz}.

For a channel described by binding energy $\epsilon_{SLJ}^{s_1s_2}$ and reduced mass $\mu_{s_1s_2}$, the (downward) mass shift and meson-meson probability are, respectively,
\be
\Delta M_{SLJ}^{s_1s_2l}	=C_{SLJ}^{s_1s_2l}\int dp \frac{p^2 |A_l(p)|^2}{\epsilon_{SLJ}^{s_1s_2}+{p^2}/{2\mu_{s_1s_2}}},
\qquad P_{SLJ}^{s_1s_2l}		=C_{SLJ}^{s_1s_2l}\int dp \frac{p^2 |A_l(p)|^2}{\left(\epsilon_{SLJ}^{s_1s_2}+{p^2}/{2\mu_{s_1s_2}}\right)^2}.
\label{massshift}
\ee
Introducing a quantity $X_{SLJ}^{s_1s_2}$, which parameterizes the reduced mass and binding energy of a given channel in terms of the spin-averaged values $\mu$ and $\epsilon$ (those corresponding to setting all spin splittings to zero),
\be
\mu_{s_1s_2}\epsilon_{SLJ}^{s_1s_2}=\mu\epsilon(1+X_{SLJ}^{s_1s_2}),
\ee
the mass shift can be expressed in a power series expansion,
\be
\Delta M_{SLJ}^{s_1s_2l}= 
C_{SLJ}^{s_1s_2l}\frac{\mu_{s_1s_2}}{\mu}\frac{1}{\epsilon} 
 \sum_{n=0}^\infty(-X_{SLJ}^{s_1s_2})^n\int dp \frac{p^2 |A_l(p)|^2}{(1+{p^2}/{2\mu\epsilon})^{n+1}}.
\label{expansion}
\ee
The first two terms in the series involve the integrals in the two equations \rf{massshift}, but with $\mu$ and $\epsilon$ in place of $\mu_{s_1s_2}$ and $\epsilon_{SLJ}^{s_1s_2}$. These terms can be thought of as the the spin-averaged mass shift and meson-meson probability. Calling these $\Delta M^l$ and $P^l$, the approximate formula for the mass shift due to a given channel is
\be
\Delta M_{SLJ}^{s_1s_2l}\approx C_{SLJ}^{s_1s_2l}\frac{\mu_{s_1s_2}}{\mu}\left(\Delta M^l-X_{SLJ}^{s_1s_2}\epsilon P^l\right),
\label{linearexpansion}
\ee
which turns out to be a reasonable approximation for $\cc$ and an excellent approximation for $\bb$. The total mass shift is the sum over those of the different channels, which is straightforward using the properties of the coefficients $C_{SLJ}^{s_1s_2l}$. The correction \rf{indhyp} to the hyperfine splitting due to channel coupling follows immediately; everything cancels except a term proportional to $\Delta_s$,
\be
-\frac{1}{9}\left(\Delta M_{\an 3P0}+3\Delta M_{\an 3P1}+5\Delta M_{\an 3P2}\right)+\Delta M_{\an 1\uP1}=-\Delta_s\sum_l P^l.
\ee
Thus to this order, in the nonrelativistic limit ($\Delta_s=0$) the result of zero hyperfine splitting survives corrections due to unquenching.  The small hyperfine splittings in Table \ref{shiftstable} are indicative of the magnitude of quadratic corrections to the expansion \rf{linearexpansion}, and the smallness of $X_{SLJ}^{s_1s_2}$ explains why the mechanism works even better for $\bb$ than $\cc$.

The mechanism relies on the factorisation of quark spin and spatial degrees of freedom and the assumption that the coupling involves the creation of a quark pair with spin one. The observed small hyperfine splittings thus supports this picture, which is also consistent with lattice QCD \cite{Burns:2006wz}. 

The same mechanism protects the hyperfine splitting of D-wave and higher $L$ mesons; thus one can predict the mass of the $\an 1D2$ bottomonium in terms of the $\an 3D{1,2,3}$, as in ref. \cite{Burns:2010qq}.

Notice in Table \ref{shiftstable} that in each model the induced hyperfine splitting is less than (or equal to) zero. If the physical hyperfine splitting is positive, as is favoured by the bulk of experimental and lattice data, then in the absence of some other effect the bare potential model splitting $\Delta_s$ must be positive. This may help to distinguish among different models, which disagree on the sign of $\Delta_s$ \cite{Godfrey:2002rp}. 

Another common feature is the hierarchy of mass splittings,
\be
\Delta M_{\an 3P2}>\Delta M_{\an 1\uP1}>\Delta M_{\an 3P1}>\Delta M_{\an 3P0},
\ee
which implies that unquenching brings meson masses closer together with respect to their bare values. Comparison of quenched and unquenched lattice QCD calculations would be an interesting test of this effect.

}  
